\documentclass[doublecol]{epl2}
\usepackage[]{graphicx}
\usepackage[]{color}
\usepackage{amsmath}
\usepackage{amssymb}
\usepackage{bm}

\newcommand{\kap}{\boldsymbol{\kappa}}
\newcommand{\rb}{{\bf r}}

\newcommand{\bp}{\boldsymbol{\partial}}
\newcommand{\vct}[1]{\mathbf{#1}}

\newcommand{\rd}{{\bf r}}

\title{
Controlling colloidal sedimentation using time dependent shear
}
\shorttitle{
Controlling colloidal sedimentation using time dependent shear
}

\date{\today}
\author{Matthias Kr\"uger\inst{1}\thanks{kruegerm@mit.edu} \and Joseph M. Brader\inst{2}\thanks{joseph.brader@unifr.ch}}
\shortauthor{M.  Kr\"uger and J. M. Brader}
\institute{
  \inst{1} Massachusetts Institute of Technology, Department of
  Physics, Cambridge, Massachusetts 02139, USA\\
\inst{2} Department of Physics, University of Fribourg, CH-1700 Fribourg, Switzerland\\
}

\abstract{
Employing a recently developed dynamical density functional theory we study the response of a colloidal sediment above a wall to shear, demonstrating the time dependent changes of the density distribution and its center-of-mass after switching shear either on or off and under oscillatory shear. Following the onset of steady shear we identify two dynamical mechanisms, distinguished by their timescales. Shortly after the onset, a transient enhancement of the packing structure at the wall reflects the self-organization into lanes. On a much longer timescale these effects are transmitted to the bulk, leading to migration away from the wall and an increase in the center-of-mass. Under oscillatory shear flow the center-of-mass enters a stationary state, reminiscent of a driven damped oscillator.
}


\begin{document}

\maketitle
Colloidal suspensions under shear have been the subject of intense research during the last decades, both theoretically as well as in simulations and experiments \cite{dhont,Larson,Bergenholtz01,joe_review}.  
Dense systems have attracted particular attention, as particle interactions can transmit the effects of shear 
into directions perpendicular to the flow, leading to nontrivial collective dynamics. 
An important consequence of applying steady shear is that the intrinsic slow dynamics exhibited by quiescent 
dense systems (and characterized by observables such as density correlators or mean squared displacements \cite{Varnik06,Besseling07,fabian_proc,Kruger11,Fuchs08b}) can be sped up drastically in all spatial 
directions, provided the shear rate exceeds the inverse of the structural relaxation time \cite{pnas}. 
The dynamical response becomes richer still when considering time-dependent shear fields for which transient 
dynamics can play a dominant role (e.g. oscillatory shear \cite{Brader_osc}). 

In real colloidal systems the buoyant mass of the suspended particles is often nonzero, such that gravity drives the 
formation of a nonuniform density distribution: A colloidal sediment (or creaming profile, in the event that the buoyant mass is negative). 
The study of colloidal sedimentation has a long history, effectively starting with the pioneering 
experimental work of Perrin 
at the start of the 19th century.   
While early theoretical works focused upon the influence of hydrodynamic interactions on the sedimentation velocity in 
dilute systems (see e.g. \cite{Batchelor82,Brady88,Vesaratchanon07}), the dynamic settling of denser suspensions has only recently been studied in detail in experiments, Brownian dynamics simulations, and theoretically, using dynamic density functional theory (DDFT) \cite{Schmidt04, Royall07,Schmidt08}. 
Incorporating hydrodynamic interactions into theories of dense suspensions is a difficult task, although 
recently some progress has been made \cite{rex2,rauscher_hydro}. 

Experiments in which shear flow has been applied to particle sediments 
(where in the early literature, mostly large, nearly non-Brownian particles were considered) 
report changes in the viscosity, attributed to a shear induced modification of the 
underlying microstructure \cite{Acrivos93}. 
Moreover, an increase in the height of the sediment has been observed with increasing flow rate. 
These two observations imply that 
shear flow acts to reduce the density within the sediment and redisperse the particles. 
This robust physical effect, now commonly referred to as `viscous resuspension', has been addressed 
in a number of studies \cite{Gadala79,Leighton86,Chapman91,Acrivos93,Shauly00} and has clear relevance 
for commercial and industrial systems in which unwanted sedimentation effects can be suppressed by  
mechanical means.

In this paper, we use the framework of DDFT to study the interplay of sedimentation and shear, 
concentrating on situations for which the direction of shear flow and the gravitational force are perpendicular. 
Standard DDFT \cite{rauscher1} for flow advection does not capture the effects of shear in this situation. 
It lacks the essential physical ingredient; namely the transmission of shear effects into directions perpendicular to the flow. We therefore employ a recently suggested 
modification of the theory which corrects this failing \cite{joe_matze1}. 
By considering the time dependence of the particle distribution we obtain insight into the microscopic mechanisms 
at work during the process of viscous resuspension. 

Following the onset of shear we observe a two step scenario. 
On a short timescale, particles close to the wall build lanes to allow for lateral movements following the shear flow. 
This local self-organization leads to a temporary reduction in the entropy of the system and drives the slow migration 
of particles to regions far away from the wall. 
Only on a much longer timescale do we observe an increase of the center-of-mass (COM) and, therefore, in the 
gravitational potential energy, leading eventually to a steady state. 
If the shear field is then suddenly switched off, we find that the equilibration dynamics show an interesting 
symmetry with that following switch on, at least for moderate shear rates. 
In the case of oscillatory shear the transient regime occurring in response to switching on the flow is followed by 
a periodic stationary state, out of phase with the applied strain.  
For large frequency the COM converges to a time independent value above the equilibrium value. 
Given the possibility of precise real time particle observation and tracking in shear cells, e.g. using confocal microscopy \cite{zausch}, we hope that our theoretical findings will motivate further experimental studies of the dynamics in 
sediments subject to shear flow.

Consider a sediment of $N$ spherical colloidal particles, $i=1\dots N$, of diameter $d$,  
dispersed in an incompressible Newtonian solvent.  
The time evolution of the probability distribution of particle positions, 
$\Psi(t)\equiv\Psi(\{ {\bf r}_i\},t)$, is given by the Smoluchowski equation \cite{dhont}
\begin{eqnarray}
\hspace*{1.1cm}\frac{\partial \Psi(t)}{\partial t} + 
\sum_{i} \bp_i\cdot {\bf j}_i =0,
\label{smol_hydro}
\end{eqnarray} 
where the probability flux of particle $i$ is given by
\begin{eqnarray}
{\bf j}_i=\big[ \kap(t)\cdot\rb_i
- D_0\cdot(\bp_i - \beta\,{\bf F}_i) \big]\Psi(t),
\label{flux}
\end{eqnarray}
where $\beta=(k_BT)^{-1}$ is the inverse thermal energy and $D_0$ is the (bare) one particle diffusion coefficient. $\kap(t)$  is the velocity gradient tensor, which we choose to be $\dot\gamma(t)\hat{\vct{x}}\hat{\vct{y}}$, i.e., we consider time dependent simple shear pointing in $x$ direction and varying in $y$ direction. Introducing dimensionless units gives space, energy and time in terms of $d$, $k_BT$ and $d^2/D_0$, respectively. The shear rate $\dot\gamma$ is in these units equal to the Peclet number Pe$=\dot\gamma d^2/D_0$.

The force on a given particle ${\bf F}_j=-\bp_j U$ is generated from the total potential $U$, made up of a pairwise additive interaction $\phi$ and an external potential $V^{\rm ext}$,
\begin{eqnarray}
U(\{ {\bf r}_i\},t)
=\sum_{i}V^{\rm ext}(\{\rb_i\},t) + \sum_{i<j}\phi(|\rb_i-\rb_j|).
\label{potential}
\end{eqnarray}
The sediment of colloidal particles shall be confined to  $y>0$, i.e., we introduce a repulsive wall in the plane $y=0$. Gravity acts perpendicular to the wall, i.e.,  in direction -$\vct{e}_y$,  and the external potential  
\begin{equation}
V^{\rm ext}(\{\rb_i\},t)=V^{\rm ext}(\{y_i\})=\frac{e^{-s(y_i-\frac{1}{2})}}{(y_i-\frac{1}{2})}+mgy_i\label{eq:pot}
\end{equation}
depends on the $y$ component of the particle positions only. The Yukawa potential (shifted by -1/2, as the center of the spheres is restricted to $y>1/2$) carries a parameter $s$ controlling the softness of the repulsion, and $mgy_i$ is the gravitational potential of particle $i$ with buoyant mass $m$. The pair potential $\phi$ is taken to be a hard sphere potential, which well describes the interaction between certain colloids,
\begin{equation}
\phi(r)=\left\{\begin{array}{cc}
0&r>{1}\\
\infty&r<{1}
\end{array}\right..
\end{equation} 
The probability flux (\ref{flux}) neglects hydrodynamic interactions and assumes that the velocity 
gradient tensor $\kap(t)$ is translationally invariant in space.  We refer the reader to a detailed discussion of these approximations in Ref.~\cite{joe_review}.

We use dynamic density functional theory (DDFT) to calculate the one-body density profile, which arises from  $\Psi$ by integration over $N-1$ particle positions \cite{dhont}
\begin{equation}
\rho(\rd,t)=N\int d^3r_2\dots d^3r_N \Psi(\{ {\bf r}_i\},t).
\end{equation} 
For the potential in Eq.~\eqref{potential}, one has $\rho(\rd,t)=\rho(y,t)$ \cite{joe_matze1}.
As discussed in detail in Ref.~\cite{joe_matze1}, standard flow advected DDFT \cite{rauscher1} does not capture the effects of shear in the considered setup, because the flow advection term vanishes exactly: It does not transmit the effects of shear, pointing in $x$ direction, to particle motion against the external potential in $y$ direction. It is therefore necessary to introduce a modified version of the DDFT equation for sheared systems \cite{joe_matze1}
\begin{equation}
\frac{\partial \rho(y,t)}{\partial t} \!=\! 
\frac{\partial}{\partial y}\Bigg[\!-\!\rho(y,t)v^{\rm fl}_y(y,t) 
+\rho(y,t)
\frac{\partial}{\partial y}\frac{\delta \mathcal{F}[\rho(y,t)]}{\delta \rho(y,t)} \Bigg].
\label{adv_ddft_mod}
\end{equation} 
The first term on the right hand side captures the velocity of particles in the $y$ direction due to shear in a mean field manner. 
As a true microscopic derivation of the shear term is still lacking, it must be viewed as an {\em ad hoc}, albeit physically motivated, addition to the theory \cite{joe_matze1}, 
\begin{equation}
v^{\rm fl}_y(y,t)=\int_{-\infty}^{\,\infty}dy' \rho(y',t)\,\bar v^k_y(y-y',t).\label{eq:mf}
\end{equation}
Eq.~\eqref{eq:mf} is a convolution of the density with a velocity $\bar v^k_y$, describing the collision process of two particles with separation $y-y'$. Within the simplest approximation it contains the equilibrium pair correlation function  $g_{\rm eq}(d)$, 
\begin{align}
\label{kernel} \bar v^k_y(\Delta y,t)=-\frac{|\dot\gamma(t)|}{\pi}\, g_{\rm eq}(d) \left(\frac{\Delta y}{d}\right)^2 \sqrt{d^2-\Delta y^2}.
\end{align}
Self-consistent updating of $g_{\rm eq}(d)$ to make it shear dependent as well as incorporation of hydrodynamic interactions are possible, in principle.  The last term in (\ref{adv_ddft_mod}) is derived from the  Helmholtz free energy  
\begin{align}
&\notag\mathcal{F}[\rho(y,t)]=\mathcal{F}^{\rm ex}[\rho(y,t)]\\ &+ \int d^3r \rho(y,t)\left[\ln (\rho(y,t)\Lambda^3) -1 +{V^{\rm ext} (y)}\right]\label{eq:F},
\end{align}
where the excess free energy $\mathcal{F}^{\rm ex}$ depends on the interaction potential $\phi$. For our hard sphere system, we use the well known Rosenfeld approximation \cite{rosenfeld}. $\Lambda$ is the thermal de Broglie wavelength. 
The starting point for our dynamical calculations is an initial equilibrium sedimentation profile with adsorption 
$\Gamma=3.7864$ (equal to the total number of particles in a vertical column of unit cross sectional area)  
calculated using the parameters $s=5$ and $mg=0.7$. 
The average particle number $N$ (albeit formally infinite as there are no boundaries in $x$ and $z$ directions) is fixed.

\begin{figure}
\includegraphics[width=8.cm,angle=0]{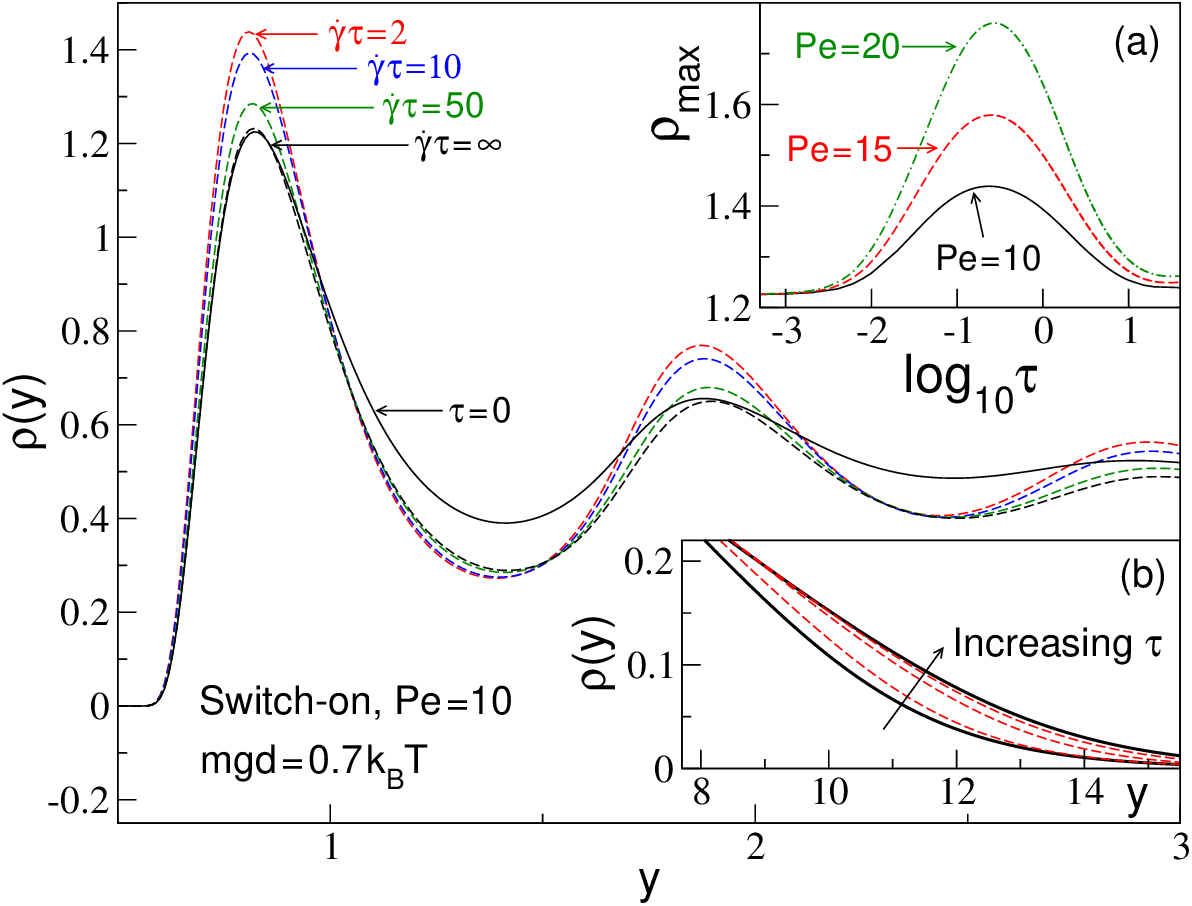}
\caption{
Evolution of a colloidal sediment above a Yukawa wall following 
the onset of steady shear flow ($\dot\gamma=10$).  
Inset (a) shows the height of the first 
peak as a function of time for $\dot\gamma=10$ (considered in the main figure) $\dot\gamma=15$ and $\dot\gamma=20$.  
Inset (b) shows the long range decay of the profile at $\tau=0$ and $\infty$ (full black 
curves), and at intermediate times $\tau=1$, $5$ and $10$ (broken red curves).
}
\label{figure1}
\end{figure}

We first consider discontinuously changing the shear rate at $t=t_{\rm on}$ from zero to a constant 
value $\dot\gamma=10$ \cite{zausch,Kruger10}, denoting $\tau=t-t_{\rm on}$. Fig.\ref{figure1} shows the corresponding time evolution of the density.
Immediately following the onset of shear, the density increases at the peaks and reduces between the peaks, i.e., the peaks sharpen: This laning in the vicinity of the wall minimizes collisions and allows the particles to move past each other. 
It is only on much longer times that the height of the first peak reduces and the tail of the profile grows monotonically to 
larger $y$ values, saturating to the steady state solution for strain $\dot\gamma \tau=\mathcal{O}(100)$. 

Inset (a) shows the nonmonotonic behavior of the height of the first peak (all other peaks evolve in a similar fashion) 
as a function of time, both for the shear rate considered in the main figure ($\dot\gamma=10$), as well as for two 
additional higher rates, $\dot\gamma=15$ and $20$. 
The time at which the maximum of the first peak overshoot occurs is not a sensitive function 
of shear rate (maxima occur at $\tau=0.240, 0.251$ and $0.272$ for $\dot\gamma=10, 15$ and $20$, respectively).  For the cases shown, the maximum height of the overshoot increases slightly faster than linearly with $\dot\gamma$. 
Inset (b) focuses on the tail of the profile, which gradually increases with time, reflecting the slow collective 
diffusion of particles away from the wall. During this slow process, the curves in inset a) approach their final values close to the equilibrium value. The general similarity of the first peak for $\tau=0$ and $\tau\to\infty$, both in height and form, is related to the fact that the overall weight of particles supported by the wall 
is equal in both situations (An analogous calculation for a strict hard-wall external field yields identical values for the contact value $\rho(R,\tau\leq0)=\rho(R,\tau\to\infty)$ \cite{joe_matze1}). 

A clear difference between steady and equilibrium profiles is seen in the vicinity of the wall ($1\lesssim y\lesssim 3$), where the former is clearly lower. 
This is a signature of particles having moved away from the wall into the tail of the distribution by collision induced diffusion (the total particle number is conserved). 

\begin{figure}
\includegraphics[width=8.2cm,angle=0]{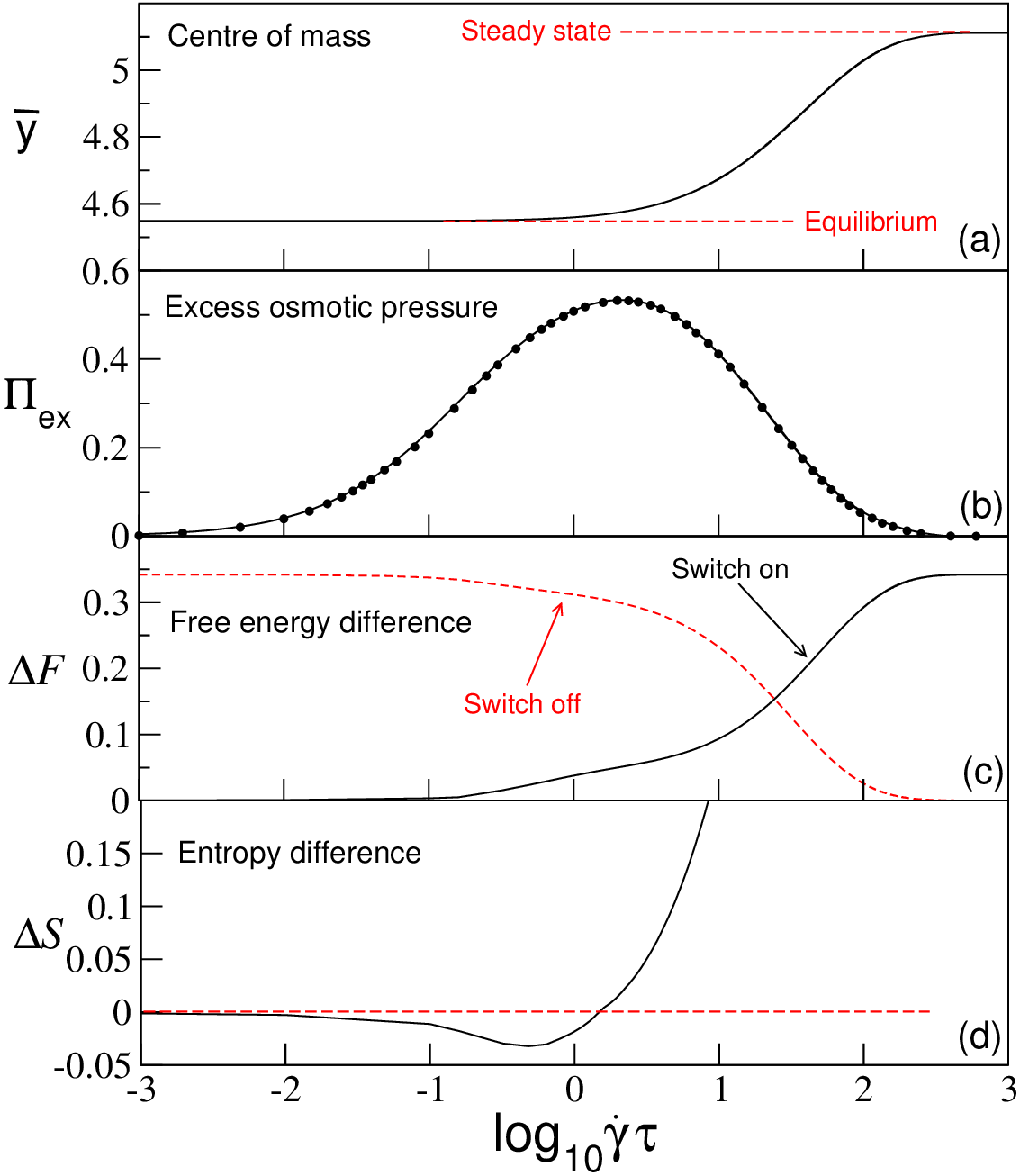}
\caption{
(a) The time development of the COM following the onset 
of steady shear flow for $\dot\gamma=10$. Saturation to a constant value 
requires strains $\dot\gamma \tau$ in excess of $100$.
(b) Excess osmotic pressure acting on the wall. 
The line is the result calculated from Eq. (\ref{pressure_wall}), after  subtraction of the 
equilibrium value $\Pi_{\rm eq}$. The circles are calculated from the 
centre-of-mass using Eq.(\ref{excess_osmotic}).  
The osmotic pressure has a maximum at a strain $\mathcal{O}(1)$.
(c) The free energy gain of the system as a function of strain. The dashed line shows the loss of free energy after switch off, see main text. 
(d) The entropy change as a function of strain. The dashed line is $\Delta S=0$ and serves as a guide to the eye.
For all curves we employ the same parameters as used in Fig.\ref{figure1}.
}
\label{figure2}
\end{figure}

The "lifting" of the density distribution against gravity after switch on can be quantified by the COM of the density distribution 
\begin{eqnarray}\label{com}
\bar{y}(t)=\frac{\int_{-\infty}^\infty \!dy\, y\, 
\rho(y,t)}{\int_{-\infty}^\infty \!dy\, \rho(y,t)}.
\end{eqnarray}
Fig.\ref{figure2}a shows the COM as a function of time after switch-on. We observe the expected monotonic increase, starting at strains greater than unity and saturating to a steady state value at strains $\sim 300$. In \cite{joe_matze1} we demonstrated that the final value of the COM increases with shear rate. The change of the density near the wall in Fig.~\ref{figure1} leads to a change of the osmotic pressure acting on the wall, given by
\begin{equation}
\Pi_{\rm w}(t)=-\int_{-\infty}^{\infty}\!dy' \rho(y',t)\frac{dV^{\rm wall}_{\rm ext}(y')}{dy'}
\label{pressure_wall}
\end{equation} 
(which reduces to the familiar sum-rule $\beta \Pi=\rho(0^+)$ for the case of a hard wall). 
For both $\tau<0$ and $\tau\to\infty$, $\Pi_{\rm w}(t)$ equals the equilibrium pressure 
$\Pi_{eq}=mg\,\Gamma$. 
In Fig.~\ref{figure2}b we show the excess pressure $\Pi_{\rm ex}(t)\equiv\Pi_{\rm w}(t)-\Pi_{\rm eq}$ as a 
function of time. 
This excess pressure lifts the particles against the Stokesian friction and must equal the average 
drag force. In the figure, we verify numerically that
\begin{equation}
\Pi_{\rm ex}(t)
\stackrel{!}{=}\frac{\Gamma}{\mu}\frac{d\bar{y}(t)}{dt},
\label{excess_osmotic}
\end{equation}
with $\mu$ the mobility (included here explicitly, despite our dimensionless description) is fulfilled at all times. 

Noting the logarithmic timescale, the excess pressure also shows the range of timescales involved, 
reaching a maximum at $\tau=0.213$ (which roughly coincides with the maximum of the first peak in Fig.~\ref{figure1}, $\dot\gamma \tau=2.4$), 
before decaying over a much longer range of strain
($\dot\gamma \tau<1000$), corresponding to the timescale where the particles migrate away
from the wall. 
The overdamped character of the colloidal dynamics is reflected by the absence of
oscillations in the pressure as a function of time.

The lifting of the COM after switch on changes the free energy, Eq.~\eqref{eq:F}, of the system, 
which, as the equilibrium marks its minimum, has to increase. 
Its increase $\Delta\mathcal{F}=\Delta U-T\Delta S=\mathcal{F}(\tau)-\mathcal{F}(\tau=0)$, as shown in  
Fig.~\ref{figure2}c, is nevertheless much smaller than the increase in potential energy $V^{\rm ext}$, because the entropic energy $-ST$, given by the remaining terms in Eq.~\eqref{eq:F}, decreases. $\Delta\mathcal{F}$ shows most clearly the two timescales and the physical mechanisms at work: The initial increase up to roughly $\dot\gamma\tau=1$ is due to the laning of particles near the wall, as observed in Fig.~\ref{figure1}. Up to this point, the COM has hardly changed (compare with Fig.~\ref{figure2}a), and this initial increase is mostly due to decreasing entropy (see Fig.~\ref{figure2}d). 
The entropy serves now as a driving force for the much slower migration of particles connected to the second increase in $\Delta\mathcal{F}$ to the final value, during which the potential energy increases and the entropic energy decreases, such that eventually $\Delta S(\tau\to\infty)=1.15$ is positive.

We note that after switch on of shear, energy is dissipated by two mechanisms: The 
vertical particle motion (in $y$-direction), and the constant dissipation of energy due to the shear 
friction, connected to the shear stress, which persists even in the steady state. 
While the former can be derived using the particle currents (see Eq.~\eqref{eq:Wdiss} below), the latter 
is at present not accessible in our theory, and we leave a full discussion of dissipation after switch 
on to future work.

Given that the steady state has been reached, we now switch off the shear flow instantaneously at $t=t_{\rm off}$ (denoting $\tilde\tau=t-t_{\rm off}$) and investigate the relaxation back to equilibrium. 
Fig.~\ref{figure2}c shows the subsequent decrease of the free energy back to the equilibrium value, which has to be dissipated via friction. The latter can be derived from the current of the one body density $j(\rd,t)= \rho(y,t)\frac{\partial}{\partial y}\frac{\delta \mathcal{F}[\rho(y,t)]}{\delta \rho(y,t)} $ in Eq.~\eqref{adv_ddft_mod}. Via Stokes' friction law, the force giving rise to this current is proportional to $j$, such that the following form arises for the energy dissipated after switch off
\begin{equation}
W_{\rm diss}(\tilde\tau)=\int d^3r\int_{0}^{\tilde\tau} d\tilde\tau' \frac{j(\rd,\tilde\tau')^2}{\mu \rho(y,\tilde\tau')}\stackrel{!}{=} \mathcal{F}(\tilde\tau=0)-\mathcal{F}(\tilde\tau).\label{eq:Wdiss}
\end{equation}
The last equality can be shown analytically using the definition in Eq.~\eqref{eq:F} and there is no need for numerical verification. Note that there is a difference between the current $j$ and the average velocity $\frac{d\bar{y}(t)}{dt}$ in \eqref{excess_osmotic}: $j$ depends on position, and can be nonzero (due to relative motion) even if $\frac{d\bar{y}(t)}{dt}$ is zero.

We note that $\Delta S(\tilde\tau\to\infty)<0$, as is typical for sedimentation: The entropic energy $-ST$ can be seen as a spring repelling the particles (or the COM) from the wall. As the COM sinks to lower values, the energy stored in the entropic spring naturally increases.

\begin{figure}
\includegraphics[width=8.cm,angle=0]{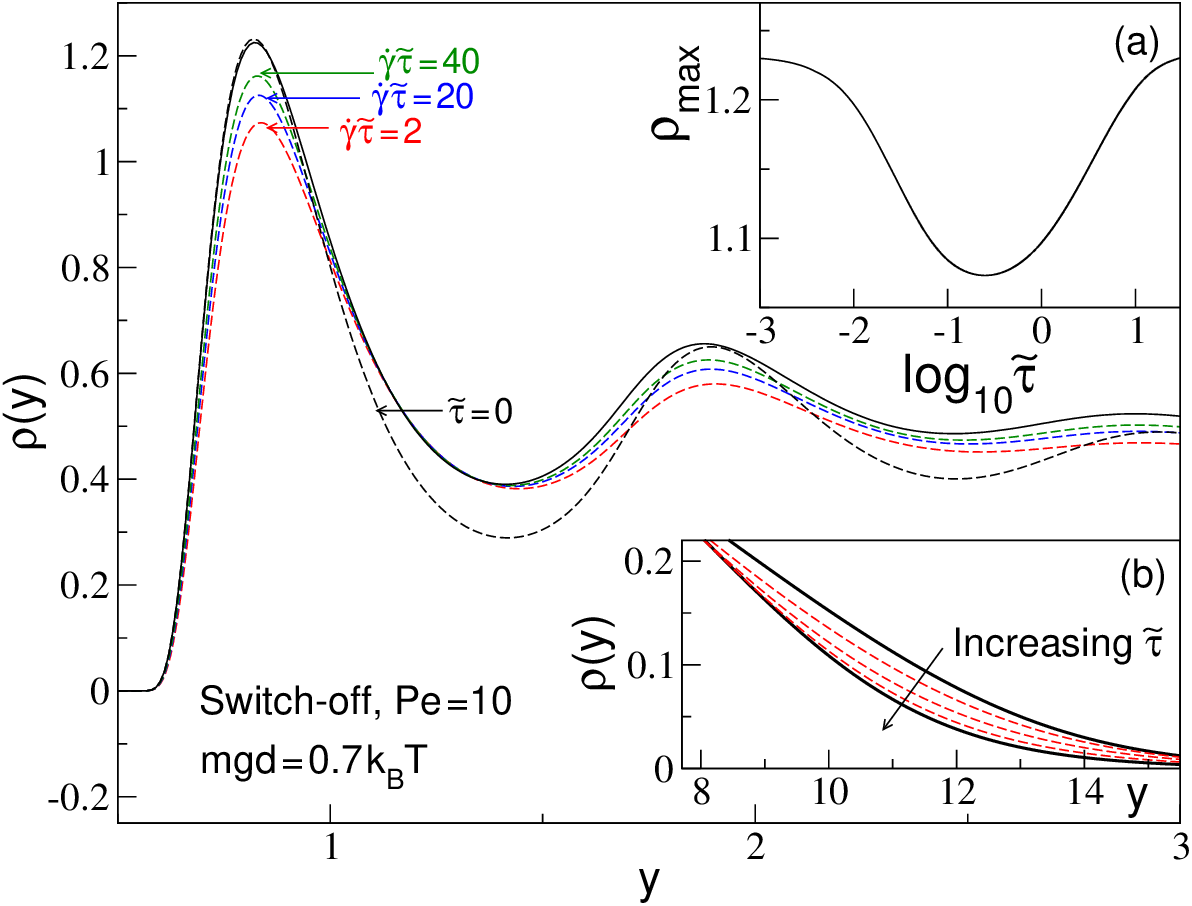}
\caption{
Analogue to Fig.\ref{figure1} for shear cessation. The initial steady state ($\dot\gamma=10$)
flow is switched off at $\tilde\tau=0$, leading to a rapid transient `undershoot' of the first
peak followed by a slower approach to equilibrium.
Inset (a) shows the height of the first peak as a function of time (in Brownian units).
A minimum occurs for $\tilde\tau=0.24$.
Inset (b) shows the long range decay of the profile at $\tilde\tau=0$ and $\infty$ (full black
curves), and at intermediate times $\tilde\tau=2$, $5$ and $10$ (broken red curves).
}
\label{figure3}
\end{figure}

In Fig.\ref{figure3} we focus on the time evolution of the density profile following switch off. 
For times shortly after cessation the heights of the peaks decrease (the main peak has its minimum value at $\tilde\tau=0.24$ (see inset (a)), accompanied with a reduction in the depth of the valleys, i.e., the peaks initially broaden, describing the inverse mechanisms to the laning after switch on. This leads to the initial decrease in $\mathcal{F}$ seen in Fig.~\ref{figure2}, because $S$ initially increases. At long times the tail of the density distribution evolves monotonically, but now from the steady state back to equilibrium. 

A symmetry in the time evolution after switch on and off is apparent in our graphs. This can be understood by linearizing the density with respect to $\dot\gamma$ after switch on and off,
\begin{align}
\rho(y,\tau)&=\rho^{\rm eq}(y)+\dot\gamma\Delta\rho(y,\tau),\\
\rho(y,\tilde\tau)&=\rho^{\rm eq}(y)+\dot\gamma\Delta\rho(y,\tau\to\infty)-\dot\gamma\widetilde{\Delta\rho}(y,\tilde\tau).
\end{align}
Of course, we have $\widetilde{\Delta\rho}(y,\infty)= \Delta\rho(y,\infty)$ such that the density approaches the equilibrium distribution after switch off. The straight forward linearization of Eq.~\eqref{adv_ddft_mod} with respect to $\dot\gamma$ yields the equality $\Delta\rho(y,\tau)=\widetilde{\Delta\rho}(y,\tilde\tau)$,
stating the symmetry of the time evolution after switch on and off. From the appearance of this symmetry in our numerical results (compare the main panels if Figs.~\ref{figure1} and \ref{figure3} as well as Fig.~\ref{figure2}c) we conclude that our chosen shear rate (despite $\dot\gamma=10\gg1$) is still not far away from this linear regime. We note that this symmetry does not depend on the details of the flow kernel in Eq.~\eqref{adv_ddft_mod}, and we expect it to be observable in experiments. However, as time grows, the full solution for finite shear rates deviates more and more from the linearized result, such that at long times, the density behaves in a more nontrivial way, leading to relaxation times which are not independent of shear rate (in contrast to the solutions for $\Delta\rho(y,\tau)$).

Let us finally consider smooth variations in shear rate by applying oscillatory shear. By standard convention the strain in an oscillatory shear experiment is given by 
\begin{equation}
\gamma(\tau)=\gamma_0\sin(\omega\tau)\Theta(\tau),
\label{strain} 
\end{equation}
where $\gamma_0$ is the strain amplitude, $\omega$ is the frequency and $\Theta(\tau)$ is the unit step function.
The oscillatory shear protocol \eqref{strain} can be  realized experimentally,
e.g. in  a simple parallel plate shear cell, which is particularly advantageous when seeking to 
combine rheological measurements with real space (confocal) microscopy (see e.g. \cite{zausch}). 

\begin{figure} 
\includegraphics[width=8.cm,angle=0]{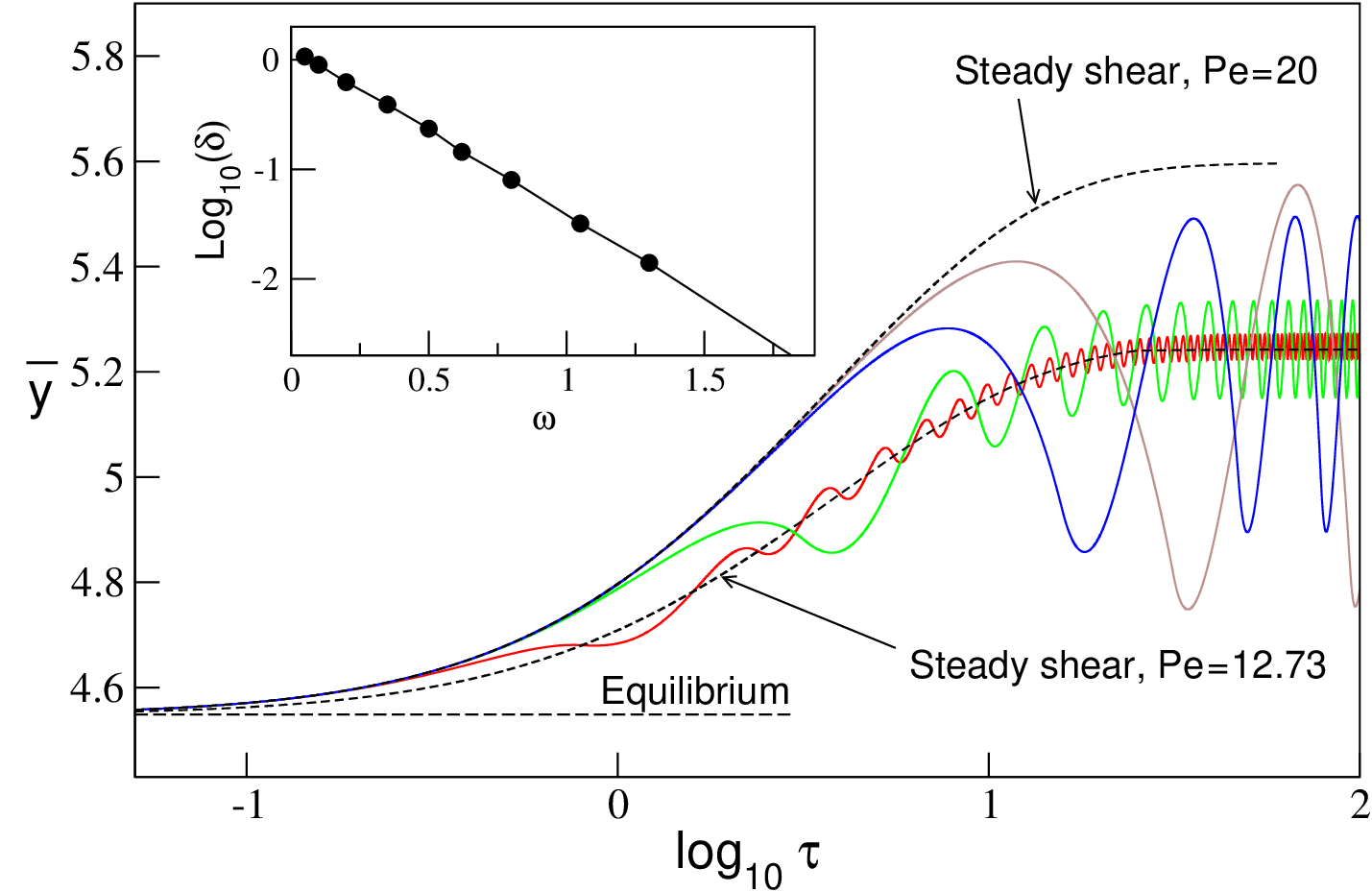}
\caption{
The time evolution of the centre-of-mass for the oscillatory shear flow in Eq.~(\ref{strain}) for frequencies $\omega=0.05$ (brown), 0.1 (blue), 0.5 (green)  and 2 (red), with fixed $\gamma_0\omega=20$. 
For long times a stationary state 
is attained. At high frequencies we sample only the average shear rate and the steady shear limit 
with $\dot\gamma=40/\pi=12.73$ is approached. 
The inset shows the phase shift of the COM relative to the applied shear strain. 
The calculated data points (circles) are consistent with an exponential decay. 
}
\label{figure4}
\end{figure}

In Fig.\ref{figure4},  we show the time evolution of the centre-of-mass for four different frequencies $\omega$ and strain amplitudes $\gamma_0$, keeping the strain rate amplitude $\gamma_0\,\omega$ (corresponding to the maximal shear rate) 
fixed at $\gamma_0\omega=20$. In all curves there is a transient dynamics  
of the COM, which initially follows the curve for steady shear with $\dot\gamma=20$, consistent with linearization of the strain rate field for short times 
$\dot\gamma(\tau)=\gamma_0\omega\cos(\omega\tau)\approx \gamma_0\omega$, until a periodic stationary state is attained. In this periodic state also the free energy (not shown) changes periodically while the system repeatedly undergoes the mechanisms described above.
For all curves shown, the stationary state is approached for times $\tau\approx 10\,\omega^{-1}$, leading to an increasingly rapid deviation from the steady shear curve for increasing $\omega$.    

The stationary $\bar{y}(t)$ resembles a driven damped oscillator with frequency $2\omega$ (the response is invariant with respect to shear direction) and a phase shift $\delta$ relative to the strain field (\ref{strain}), such that its fundamental harmonic is $\sim \sin(2\omega\tau+\delta)$. The inset of Fig.~\ref{figure4} shows $\delta$ as a function of $\omega$, again for fixed $\gamma_0\omega$. For $\omega\to0$,  $\dot\gamma$ changes slower and slower, and the COM $\bar{y}^{\rm osc}$ in oscillatory flow can be expressed by its value $\bar{y}^{\rm st}(\dot\gamma)$ in steady shear, as 
\begin{equation}
\lim_{\omega\to0,\gamma_0\omega=const.}\bar{y}^{\rm osc}(\tau)=\bar{y}^{\rm st}(\dot\gamma(\tau)).
\end{equation}
In this limit, $\delta$ thus approaches a value close to $\pi/2$, and the COM will oscillate between $\bar{y}^{\rm st}(\dot\gamma=0)$ and $\bar{y}^{\rm st}(\dot\gamma=\gamma_0\omega)$ (the equilibrium and steady shear lines in the figure). As $\omega$ increases, $\delta$ decreases (nearly exponentially), and the COM is more and more in phase with the strain. The phase shift $\delta$ is reminiscent of the well known phase shift of the stress response with respect to strain\cite{Larson,Brader_osc}, as described by the storage ($G'$) and loss ($G''$) moduli. $G'$  measures the stress in phase with distortion $\gamma$, giving the elastic response, while  $G''$  measures the stress in phase with rate $\dot\gamma$, thereby quantifying viscous losses. As mentioned above, a detailed study of dissipation in our system (including COM motion) will be addressed in future work, which appears indeed of importance for a complete description of viscous loss: If the colloids in an experiment are not perfectly density matched with the solvent (as is always the case in practice) the dissipation mechanism identified here, dragging the centre-of-mass up and down through the solvent, will provide an additional contribution to the loss modulus.

As the frequency is increased, we observe that the oscillations in $\bar{y}(\tau)$ decrease in amplitude. In this limit, the COM is too slow to follow the quick changes of the shear rate and the envelope function enclosing the oscillations converges smoothly to a finite value larger than  the equilibrium value. It is intuitive that this value corresponds to the {steady} shear result taken at the average of $|\dot\gamma|$ during one cycle, given by $\langle|\dot\gamma|\rangle=\frac{2\gamma_0\omega}{\pi}$. We thus anticipate that the COM can also in this limit be expressed by its value in steady shear, 
\begin{eqnarray}
\lim_{\omega\to\infty,\gamma_0\omega=const.}\bar{y}^{\rm osc}(\tau)=\bar{y}^{\rm st}\left(\dot\gamma=\frac{2\gamma_0\omega}{\pi}\right).\label{eq:lim}
\end{eqnarray}
In Fig.~\ref{figure4} we show the COM as function of time after switch on for $\dot\gamma=12.73$, corresponding to the average rate for $\gamma_0\omega=20$.  
The present theory obeys \eqref{eq:lim} for any fixed value of $\gamma_0\omega$. 
Physically, however, we would expect the COM to approach the equilibrium value for 
$\gamma_0\ll1$, as the particles just move affinely back and forth 
over a strain range so small that they do not collide. 
We therefore suspect that the flow kernel \eqref{kernel} only gives physically useful results for strains 
large enough that many particle collisions happen during any cycle.  
The fact that our approach can only resolve time intervals within which many collisions have occurred 
represents an intrinsic limitation of our mean-field approach. 

Regarding an experimental sample, deviations from a linear shear profile, giving rise to deviations from Eq.~\eqref{eq:lim}, could be anticipated 
when $\omega^{-1}$ becomes comparable to the time taken for shear waves in the solvent to 
diffuse across the physical sample from one bounding plate to the other.  Despite these limitations,  the behavior described by Eq.~\eqref{eq:lim} should be observable in experiments for a certain frequency range.

To conclude: Under time dependent shear the response of the density distribution of a colloidal sediment is also time dependent, as can be observed from its COM. 
After switch on of shear, a two step mechanism can be identified, consisting of laning followed by 
migration. 
Several of the quantities discussed in the present work are accessible using real time particle 
tracking \cite{zausch} and can be directly tested in experiment. 
Future work will focus on the time dependent shear stress and energy dissipation, aiming at a more 
complete description of the phenomenon of viscous resuspension.

This work was supported by the DFG grant No. Kr 3844/1-1 and the Swiss National Science 
Foundation.


\begin{thebibliography}{29}

\bibitem{dhont}
J.K.G. Dhont, \emph{An Introduction to Dynamics of Colloids} (Elsevier science,
  Amsterdam, 1996)

\bibitem{Larson}
R.G. Larson, \emph{The Structure and Rheology of Complex Fluids} (Oxford
  University Press, New York, 1999)

\bibitem{Bergenholtz01}
J.~Bergenholtz, Current Opinion in Colloid \& Interface Science \textbf{6}, 484
  (2001)

\bibitem{joe_review}
J.~Brader, J. Phys.: Condens. Matter \textbf{22}, 363101 (2010)

\bibitem{Varnik06}
F.~Varnik, J. Chem. Phys. \textbf{125}, 164514 (2006)

\bibitem{Besseling07}
R.~Besseling, E.R. Weeks, A.B. Schofield, W.C.K. Poon, Phys. Rev. Lett.
  \textbf{99}, 028301 (2007)

\bibitem{fabian_proc}
O.~Henrich, F.~Weysser, M.E. Cates, M.~Fuchs, Phil. Trans. R. Soc. A
  \textbf{367}, 5033 (2009)

\bibitem{Kruger11}
M.~Kr\"uger, F.~Weysser, M.~Fuchs, arXiv:1107.1175

\bibitem{Fuchs08b}
M.~Fuchs, Adv. Polym. Sci. \textbf{236} (2010)

\bibitem{pnas}
J.~Brader, T.~Voigtmann, M.~Fuchs, R.~Larson, M.~Cates, Proc. Natl. Acad. Sci.
  U.S.A. \textbf{106}, 15186 (2009)

\bibitem{Brader_osc}
{J.M. Brader {\em et al}}, Phys. Rev. E \textbf{82}, 061401 (2010)

\bibitem{Batchelor82}
G.K. Batchelor, Journal of Fluid Mechanics \textbf{119}, 379 (1982)

\bibitem{Brady88}
J.F. Brady, L.J. Durlofsky, Phys. Fluids \textbf{31}, 717 (1988)

\bibitem{Vesaratchanon07}
S.~Vesaratchanon, A.~Nikolov, D.T. Wasan, Advances in Colloid and Interface
  Science \textbf{134-135}, 268 (2007)

\bibitem{Schmidt04}
M.~Schmidt, M.~Dijkstra, J.P. Hansen, J. Phys.: Condens. Matter \textbf{16},
  S4185 (2004)

\bibitem{Royall07}
C.P. Royall, J.~Dzubiella, M.~Schmidt, A.~van Blaaderen, Phys. Rev. Lett.
  \textbf{98}, 188304 (2007)

\bibitem{Schmidt08}
M.~Schmidt, C.P. Royall, A.~van Blaaderen, A.~Dzubiella, J. Phys.: Condens.
  Matter \textbf{20}, 494222 (2008)

\bibitem{rex2}
M.~Rex, H.~L\"owen, Eur. Phys. J. E \textbf{28}, 139 (2009)

\bibitem{rauscher_hydro}
M.~Rauscher, J. Phys.: Condens. Matter \textbf{22} (2010)

\bibitem{Acrivos93}
A.~Acrivos, R.~Mauro, X.~Fan, Int. J. of Multiphase Flow \textbf{19}, 797
  (1993)

\bibitem{Gadala79}
F.A. Gadala-Maria (1979), ph.D. Thesis, Standford Univ., Calif.

\bibitem{Leighton86}
D.~Leighton, A.~Acrivos, Chem. Eng. Sci. \textbf{41}, 1377 (1986)

\bibitem{Chapman91}
B.K. Chapman, D.T.J. Leighton, Int. J. Multiphase Flow \textbf{17}, 469 (1991)

\bibitem{Shauly00}
A.~Shauly, A.~Wachs, A.~Nir, Int. J. of Multiphase Flow \textbf{26}, 1 (2000)

\bibitem{rauscher1}
M.~Rauscher, A.~Dominguez, M.~Kr\"uger, F.~Penna, J. Chem. Phys. \textbf{127},
  244906 (2007)

\bibitem{joe_matze1}
J.~Brader, M.~Kr\"uger, Mol. Phys. \textbf{109}, 1029 (2011)

\bibitem{zausch}
J.~Zausch, J.~Horbach, M.~Laurati, S.~Egelhaaf, J.~Brader, T.~Voigtmann,
  M.~Fuchs, J. Phys.: Condens. Matter \textbf{20}, 404210 (2008)

\bibitem{rosenfeld}
Y.~Rosenfeld, Phys. Rev. Lett. \textbf{63}, 980 (1989)

\bibitem{Kruger10}
M.~Kr\"uger, F.~Weysser, T.~Voigtmann, Phys. Rev. E \textbf{81}, 061501 (2010)

\end{thebibliography}
\end{document}